\documentclass[aps,prl,twocolumn,amsfonts,superscriptaddress,showpacs]{revtex4-1}

\usepackage{amsmath,amssymb,mathrsfs}
\usepackage{latexsym}
\usepackage{graphicx} 
\usepackage{epstopdf}
\usepackage{graphicx,epstopdf,color}
\usepackage{amsfonts}
\usepackage{amsmath,amssymb,mathrsfs}
\usepackage{hyperref}
\usepackage{caption}
\usepackage{bm}

\newcommand{\nn}{\nonumber}

\allowdisplaybreaks[1]
\begin{document}
\title{Bosonic Mott Insulator with Meissner Currents}
\author{Alexandru Petrescu}
\affiliation{Department of Physics, Yale University, New Haven, CT 06520, USA}
\affiliation{Centre de Physique Th\' eorique, \' Ecole Polytechnique, CNRS, 91128 Palaiseau C\' edex, France}

\author{Karyn Le Hur}
\affiliation{Centre de Physique Th\' eorique, \' Ecole Polytechnique, CNRS, 91128 Palaiseau C\' edex, France}
\date{\today}

\begin{abstract}
We introduce a generic bosonic model exemplifying that (spin) Meissner currents can persist in insulating phases of matter. We consider two species of interacting bosons on a lattice. Our model exhibits separation of charge (total density) and spin (relative density): The charge sector is gapped in a bosonic Mott insulator phase with total density one, while the spin sector remains superfluid due to interspecies conversion. Coupling the spin sector to the gauge fields yields a spin Meissner effect reflecting the long-range spin superfluid coherence. We investigate the resulting phase diagram and describe other possible spin phases of matter in the Mott regime possessing chiral currents as well as a spin-density wave phase. The model presented here is realizable in Josephson junction arrays and in cold atom experiments. 
\end{abstract}

\maketitle

Interacting bosons in magnetic fields exhibit a range of interesting phenomena, from field expulsion in the Meissner-Ochsenfeld effect of superconductivity \cite{Meissner,BCS,deGennes} to the realization of topologically exotic ordered states \cite{topo}. The realization of ultracold atomic systems allows to meticulously engineer such exotic phases of matter, in particular, through the realization of synthetic gauge fields  \cite{gauge,Aidelsburger,Simonet, HofstadterCA}. The presence of multiple particle species has also been addressed \cite{sengstock_group,bloch_group,esslinger_group}. Analogous phase transitions have been studied with Josephson-junction arrays in real magnetic fields \cite{geerligs_et_al,van_der_zant_et_al,vanoudenaarden,glazman_larkin}. With respect to systems with multiple species of particles, the phenomenon of interspecies coherence has been explored in Bose-Einstein condensates \cite{Shin,Leblanc}, bilayers of dipolar Fermi gases \cite{lutchyn_et_al}, quantum Hall bilayers \cite{yang_et_al}, excitons in quantum wells \cite{san_diego} and bilayer graphene \cite{bilayergra}, polariton condensates \cite{BlochJ}. Interspecies coherence and spin-charge separation have been studied for bosons \cite{OrignacGiamarchi,donohue_giamarchi,Crepin,Kleine,Isacsson}, giving rise to a Meissner effect in the superfluid regime \cite{OrignacGiamarchi}. Similar physics has been studied with fermions  \cite{carr_et_al}. Bosonic systems with time-reversal symmetry breaking and spin-charge separation  yield rich phase diagrams \cite{GT,Klanjsek,dhar_et_al,wong_duine,zaletel_et_al}. 

In this Letter, we put such ingredients together and re-explore  the phenomenon of spin and charge separation in a two-species bosonic system \cite{Kleine} incorporating the presence of (artificial) gauge fields. 

In optical lattices, a  transition between a bosonic superfluid to a Mott insulator has been observed experimentally \cite{Greiner}, in agreement with theory \cite{Fisher,jaksch_bruder_et_al}, as well as disorder effects resulting in glassy phases \cite{Fisher,Giamarchi,glassy}. Here, we restrict ourselves to a Mott insulating regime with total density one. The system under consideration constitutes an example of a time-reversal symmetry breaking Mott phase of bosons with chiral pseudo-spin currents. A prerequisite is the phase coherence between the two species which is realized by Josephson coupling, and explicitly breaks the $U(1)$ phase symmetry. Counter-flowing spin Meissner currents with zero net charge transfer can be induced by low-flux artificial magnetic fields.  Our main result is a proof that the Meissner currents subsist as the system enters the total density Mott phase independently of the dimensionality of the system. 

We consider two species of interacting lattice bosons where the conversion term mimics the Josephson-type coupling. In a generic gauge field, the Hamiltonian reads

\begin{figure}[t]
\includegraphics[width=0.85\columnwidth]{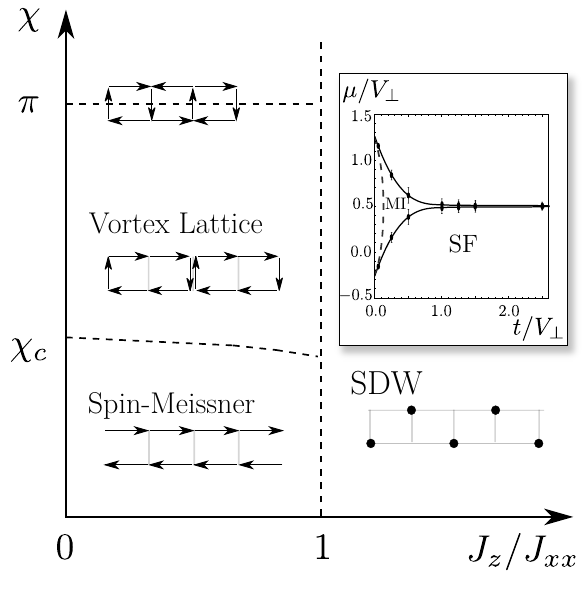}
\label{Fig:1}
\caption{Phase diagram for the effective gauged spin-$\frac{1}{2}$ model in Eq. (\ref{Eq:XXZ}) built for large repulsive terms $U$ and $V_{\perp}$. In the $XY$ limit, depending on flux, there is a spin Meissner phase or a vortex lattice phase (the direction of current patterns is shown in each phase). The inset shows the Mott lobe with total density $\rho=1$ of interest, obtained using DMRG for the one-dimensional model. The dashed line is the mean-field theory result.}
\end{figure}

\begin{eqnarray}
\label{Eq:H}
H &=& -t \sum_{\alpha, \langle i j \rangle}  e^{i a A^\alpha_{ij}}b_{\alpha i}^\dagger b_{\alpha j}  - g \sum_{\alpha, i} e^{-i a' A_{\perp i}} b^\dagger_{2 i} b_{1 i} + h.c.,  \nonumber \\
&+& \frac{U}{2} \sum_{\alpha,i} n_{\alpha i} (n_{\alpha i} - 1) + V_\perp \sum_i n_{1 i}n_{2 i} - \mu \sum_{\alpha i} n_{\alpha i}. 
\end{eqnarray}
$a A^\alpha_{ij}$ is the Peierls phase acquired by a particle of species $\alpha=1,2$, and $a' A_{\perp i}$ the phase acquired upon species conversion. Within our notations, $a$ and $a'$ depict lattice spacing in the longitudinal and transverse directions, respectively (see Fig. 1). The model in Eq. (\ref{Eq:H}) exhibits the Mott insulator to superfluid phase transition mentioned earlier. The phase boundaries can be calculated using variational mean-field theory and, for a one-dimensional lattice, exact density matrix renormalization methods \cite{kuehner_et_al} (these approaches are summarized in the Supplementary Material \cite{supplement}). The Mott insulator is unambiguously characterized by vanishing total density fluctuations. In the limit of hard-core bosons $(U\rightarrow +\infty)$, increasing either the interspecies coupling $V_\perp$ or the conversion $g$ from zero is sufficient for the existence of the Mott phase with $ \rho = 1$; the limits of Mott phase for vanishing kinetic terms are $\mu = -g$ and $\mu=V_\perp + g$.  On the superfluid side, bosons condense (quasi-condense in one dimension).  Interspecies phase coherence can still remain in the Mott phase, $\langle b_1^\dagger b_2 \rangle \neq 0$, due to the Josephson coupling. 

We define Meissner currents to satisfy the twofold condition: 1. vanishing between the species (there is no current proportional to $g$); 2. nonzero for the same species, and proportional to minus the Peierls phase acquired by a particle. The current of the relative density operator $\dot{n}_{1i} - \dot{n}_{2i}$ separates into intraspecies and interspecies components $j_\sigma = j_\parallel (i\rightarrow j) + j_\perp (i)$; these are 
\begin{eqnarray}
\label{Eq:js}
j_\parallel &=& i t ( - e^{i a A^{1}_{ij}} b_{1i}^\dagger b_{1j} + e^{i a A^{2}_{ij}} b_{2i}^\dagger b_{2j} ) + \text{H.c.}, \nonumber \\
j_\perp &=&- 2 i g b_{1i}^\dagger b_{2i} e^{i a' A_{\perp i}} + \text{H.c.} 
\end{eqnarray}
Outside the Mott lobe, the phase-angle representation is justified $b^\dagger_{1,2 i} = \sqrt{n} e^{i\theta_{1,2 i}}$ (in this reasoning, $n=\rho/2$ represents the mean (superfluid) density in each species). The conversion takes the form of a Josephson coupling 
\begin{equation}
\label{Eq:JoCo}
-g \cos(a'A_{\perp i} + \theta_{1i} - \theta_{2i}).
\end{equation} 
For strong $g$, the superfluid phases will be pinned by this term such that $a'A_{\perp i} + \theta_{1i} - \theta_{2i} = 0$. Then $j_\perp$ vanishes and furthermore in the small field limit we may expand to obtain the Meissner form of the intraspecies current
\begin{equation}
\label{Eq:jM}
\langle j_\parallel \rangle = - 2 t n \;\text{phase}_{ij}.
\end{equation} 
We have defined the phase around a plaquette, $\text{phase}_{ij} = (A_{ij}^2 - A_{ij}^1)a + (A_{\perp i} - A_{\perp j})a'$, which is invariant under a lattice gauge transform with scalars $\varphi_i^\alpha$, $A^{\alpha}_{ij}\rightarrow A^{\alpha}_{ij} + (\varphi_j^{\alpha} - \varphi_i^{\alpha})/a$ and 
$A_{\perp i}\rightarrow A_{\perp i}+(\varphi^2_{i} -\varphi^1_{i})/a'$. As expected, there is a Meissner effect in the superfluid sector in the low field limit, as checked in Ref. \cite{OrignacGiamarchi}, for example, in the specific case of one-dimensional systems. 

In fact, as we argue below, the same remains true inside the Mott phase with total density $\rho=1$. To show this, we place ourselves in the limit of large Mott gap favored by the interplay between the prominent Hubbard term $U$ and the inter-species repulsion $V_{\perp}$. In this Mott phase at $ \rho =1$, the density $\rho = (n_1 + n_2)$ is not fluctuating. The limit of strong interactions has been achieved in ultracold atoms \cite{Paredes}. A gauged spin-$\frac{1}{2}$ model is easily obtained in the limit of strong interactions, as summarized in the Supplementary Material \cite{supplement}. The two species are the Schwinger bosons in the representation of spin $\frac{\rho}{2}$ operators. The relative density corresponds to $\sigma_z=b_1^\dagger b_1 - b_2^\dagger b_2$. As demonstrated in boson language,  $\sigma_z$ fluctuates in the Mott phase. This is due to a transverse magnetic field in the $x-y$ plane, $-g \cos(a' A_{\perp i })\sigma_i^x + g \sin(a' A_{\perp i }) \sigma_i^y$. (We have used 
$\sigma^x = b_1^\dagger b_2 + h.c.$ and $\sigma_y = -i b_1^\dagger b_2 + h.c.$). 

The generic Hamiltonian for pseudospin we obtain is

\begin{eqnarray}
\label{Eq:XXZ}
H_\sigma = -\sum_{\langle i j \rangle} \left( 2J_{xx} (\sigma_i^+ \sigma_j^- e^{i a A_{ij}^\sigma} + \text{H.c.}) - J_z \sigma_z^i \sigma_z^j \right) \nonumber \\
-g \sum_i (\sigma_i^x \cos(a' A_{\perp i}) - \sigma_i^y \sin(a' A_{\perp i}) ),
\end{eqnarray}

\noindent with $J_{xx} = \frac{t^2}{V_\perp}$ and $J_z = t^2\left(-\frac{2}{U} + \frac{1}{V_\perp}\right)$, and $A^\sigma = A^1 - A^2$. Setting $V_\perp = U/2$ or $J_z=0$ yields the gapless $XY$ phase of Eq. (\ref{Eq:XXZ}) and the Heisenberg antiferromagnetic chain is reached for $U\rightarrow +\infty$. In the absence of gauge fields, the $XY$ term is ferromagnetic. For experimentally feasible values the Ising term is antiferromagnetic $(J_z>0)$. These types of spin models have been addressed in various contexts \cite{carr_et_al,Altmanetal,Kuklov,affleck_garate}.

At weak Ising interactions, the ferromagnetic $XY$ order corresponds to superpositions $\alpha b_1^\dagger + \beta b_2^\dagger$. These are just two distinct regimes for the unit density Mott phase of Fig. 1. The pseudospin current associated with $\sigma^z$ is 

\begin{eqnarray}
j_\parallel &=& 2J_{xx} \big[ \cos(A_{ij}^\sigma)(\sigma_i^y \sigma_j^x - \sigma_i^x \sigma_j^y)   \nonumber \\
&+&  \sin(A_{ij}^\sigma)(\sigma_i^x \sigma_j^x + \sigma_i^y \sigma_j^y) \big], \nonumber \\
j_\perp &=& -2 g \left[ \cos(a' A_{\perp i}) \sigma_i^y + \sin(a' A_{\perp i} )\sigma_i^x\right]. 
\end{eqnarray}

\noindent Considering the $XY$ ordered phase, we define the expectation values of spin operators in this state as $\langle \sigma_{i}^x \rangle = \cos(\Theta_{\sigma i})$ and $\langle \sigma_i^y \rangle = \sin(\Theta_{\sigma i})$ (we define $\Theta_{\sigma i}=\theta_{1i}-\theta_{2i}$). A minimization of the resulting variational energy for \textit{strong} $g$ shows that these phases are pinned $\Theta_{\sigma i} + a' A_{\perp i} = 0$. Then interspecies current $\langle j_\perp\rangle = 2 g \sin( \Theta_{ \sigma i } + a' A_{ \perp i } )$ vanishes, and a similar Meissner current to that of Eq. (\ref{Eq:jM}) is obtained, $\langle j_{\parallel}\rangle = -2J_{xx}\;\text{phase}_{ij}.$ This strong coupling form is analogous to the form of Eq. (\ref{Eq:jM}) computed in the superfluid phase, where $J_{xx}$ has replaced the kinetic term $t$. The condition of strong $g$ coupling is in fact naturally achieved via renormalization group arguments.  Associated with the spin-charge separation, there are two relevant energy scales, the Mott scale, and the scale associated with phase coherence or the Meissner effect, on which two-point correlations of $b_1^\dagger b_2$ are observable. In the strongly interacting regime $(U,V_\perp) \gg (t,g)$, the Mott energy scale is formally ``infinite'' compared to the scale of the Meissner phase. 

Considering first the one-dimensional limit, we use the technique of bosonization and a renormalization-group treatment to draw the phase diagram of the model in Eq. (\ref{Eq:XXZ}). The standard treatment is to express the spin $\frac{1}{2}$ operators in terms of fermion field operators via the Jordan-Wigner transformation \cite{Giamarchi,Haldane2,Affleck}.
The resulting free part of the Hamiltonian has dispersion $\epsilon_k = - 4 J_{xx} \cos(k a - \chi)$ and Fermi velocity $v_F = |4 a J_{xx}|$. Within our notations, the flux $\chi$ reads ``$\text   \;\text{phase}_{i i+1}$'', and the Fermi surface is delimited by 
$k_F = \pm\frac{\pi}{2a} +\frac{\chi}{a}$ for the half-filled band with an additional flux. The Ising term produces next neighbor interactions. 

The low-energy spectrum is then mapped to a continuum bosonic theory \cite{Giamarchi,Haldane2,Affleck}. Introducing fields $\phi_{\sigma},\theta_{\sigma}$ with commutator $[\nabla\theta_\sigma(x),\phi_\sigma(x')]=-i\pi\delta(x-x')$, the continuum Hamiltonian has the form

\begin{eqnarray}
\label{Eq:HSig}
H_\sigma &=& \frac{1}{2\pi}\int dx \left( u_\sigma K_\sigma  (\nabla \theta_\sigma - A^\sigma)^2 + \frac{u_\sigma}{K_\sigma} (\nabla \phi_\sigma)^2 \right) \nn \\ 
&-& \frac{2J_z}{(\pi^2 a)} \int dx \cos(4 \phi_\sigma) \\ 
&-& \frac{2 g}{\sqrt{2\pi}a} \int dx \cos\left(\theta_\sigma(x) + a' A_\perp \right) \left(1 + (-1)^\frac{x}{a}\cos 2\phi_\sigma \right) \nn.
\end{eqnarray}

\noindent  
The sine-Gordon term in Eq. (\ref{Eq:HSig}) has been approximated by keeping only $q \sim 0$ terms in the density operators. The speed of sound is $u_\sigma = v_F \left[ 1 + 16 a J_z/\pi v_F \right]^{\frac{1}{2}}$; the Luttinger parameter $K_\sigma = \left[ 1 + 16 a J_z/\pi v_F \right]^{-\frac{1}{2}}$ is a measure of interaction strength. $K_\sigma = 1$ for the $xy$ limit and decreases as antiferromagnetic $J_z>0$ is turned on. Gauge invariance can be checked simply by shifts of 
$\theta_\sigma \rightarrow \theta_\sigma + \varphi$. 

We now turn to the phase diagram in Fig. 1 for our effective model. Whenever $J_z > J_{xx}$, dominant Ising interactions induce an antiferromagnetic spin density wave and there is no (Meissner) current. The corresponding inset shows a charge density wave of the bosons $b_{1,2}$, depicted as localized in two layers. The $\phi_\sigma$-dependent sine-Gordon term is irrelevant if $K_\sigma>\frac{1}{2}$, or $J_z < J_{xx}$. The remaining sine-Gordon term is $\propto g \cos(\theta_\sigma + \chi \frac{x}{a})$, where we have chosen the Landau gauge with all flux on the conversion term. For infinitesimal flux, we may neglect the influence of $\chi$. For $K_\sigma > \frac{1}{8}$, this term flows to strong-coupling, and it is associated with the following energy gap \cite{supplement} (we define $g_{\sigma}=ga/u_{\sigma}$)

\begin{equation}
\label{Eq:Delta_sigma}
\Delta_\sigma \sim \frac{u_{\sigma}}{a} g_\sigma^{\frac{1}{2-\frac{1}{4K_\sigma}}}.
\end{equation}

\noindent This expression assumes that the bare value of $g\ll J_{xx}$. For nonzero fluxes $\chi$, the energy scale in Eq. (\ref{Eq:Delta_sigma}) defines the critical flux $\chi_c$ at which the system undergoes a transition to a vortex lattice phase of the commensurate-incommensurate type \cite{Giamarchi}. Below this critical field, the phase is the spin-Meissner low-field Mott phase, characterized by zero interspecies (or bulk) currents, and counterflowing intraspecies currents. The following correlation function $\langle \sigma^+ (x) \sigma^-(0)\rangle \sim \langle e^{-i\theta_\sigma(x)} e^{+i\theta_\sigma(0)} \rangle \sim \langle e^{-i\theta_\sigma(x)} \rangle \langle e^{i\theta_\sigma(0)} \rangle$ is asymptotically constant at large distances. This situation corresponds to $XY$ order polarized (definite $\langle \theta_\sigma \rangle$) due to the in-plane field $g$. To return to the original boson operators, $\theta_\sigma = 0$ corresponds to a ``bonding'' state produced by the operator $(b_1^\dagger + b_2^\dagger )/\sqrt{2}$. 

Above the critical field $\chi_c$, currents organize in a vortex lattice, corresponding to commensurate values of the flux \cite{OrignacGiamarchi}. A flux of $\frac{p}{q}2\pi$ corresponds to $p$ vortices in $q$ unit cells as found from the expectation value of the current operator $\langle j_\perp\rangle \propto g \sin \left( \frac{\pi}{q} + \frac{2\pi p}{q}\frac{x}{a}  \right)$. When the flux is further increased to half the elementary flux per plaquette, $\chi = \pi$, the sine-Gordon term oscillates $(-1)^\frac{x}{a} g \cos(\theta_\sigma )$ and is naively irrelevant, but at second order in perturbation theory \cite{OrignacGiamarchi} the oscillatory part disappears and the contribution is proportional to $ \frac{g^2}{u_{\sigma}} \cos(2\theta_\sigma)$. This pins the field $\theta_\sigma$ to a new minimum which gives a staggered current configuration $\langle j_\perp \rangle \propto (-1)^{\frac{x}{a}}$ as shown in Fig. 1 (horizontal line at $\chi=\pi$). This phase corresponds to the ``chiral Mott insulator'' phase of boson ladders discussed in Ref. \cite{dhar_et_al}, and which exists in fermion ladders at weak field \cite{carr_et_al}. For completeness, we have checked the precise Meissner current pattern by exact diagonalization of small systems. Each species is localized in one of two chains composing a ladder. We have considered ladders of up to $10$ rungs. We confirmed numerically the Meissner current of Eq. (\ref{Eq:jM}), at small flux, as well as the vortex lattice and staggered current configurations depicted as insets in the phase diagram of Fig. 1.

The derivation of the effective $XY$ model of Eq. (\ref{Eq:HSig}) can be extended to $(d+1)$-dimensions via a variational approach (See Supplementary Material at Ref. \cite{supplement}; to substantiate our analysis of two dimensional systems, we also consider an array of coupled ladders). Starting from Eq. (\ref{Eq:XXZ}), we introduce the following pseudospin coherent state $\left|\psi\right\rangle = \prod_i ( \cos \phi_{\sigma i} \left|\uparrow\right\rangle_i + e^{i\theta_{\sigma i}} \sin \phi_{\sigma i} \left|\downarrow\right\rangle_i )$. The azimuthal and polar angles are $2\phi_\sigma$ and $\theta_\sigma$, respectively. Expanding about a saddle point corresponding to $XY$ order, taking the continuum limit and expanding in gradients, we arrive at the following continuum Hamiltonian

\begin{eqnarray}
\label{Eq:XYdDim}
H_\sigma [ \theta_\sigma, \phi_\sigma ] = \frac{1}{2}\int \frac{d^d x}{a^{d-2}} J_{xx} \left(\nabla \theta_\sigma - A^\sigma \right)^2 \nonumber \\
 - \int \frac{d^d x}{a^d} g \cos\left(\theta_\sigma + a'A_\perp\right).
\end{eqnarray}

\noindent Firstly, if we restore the quantum character of $\theta_\sigma, \phi_\sigma$, this form is identical to the one of Eq. (\ref{Eq:HSig}) in the one-dimensional limit with $J_z$ taken to zero. In addition, the argument proving the existence of the Meissner current was independent of dimension.

Secondly, viewing Eq. (\ref{Eq:XYdDim}) as the energy of a classical two-dimensional system, the first term in Eq. (\ref{Eq:XYdDim}) can be rewritten as $\frac{1}{2}\int d^2 x \rho_\sigma (\nabla\theta_\sigma(x)-A^\sigma)^2$, where $\rho_\sigma \equiv J_{xx}$ is the pseudospin rigidity, which is accessible experimentally. This gauged $XY$ model undergoes a Berezinskii-Kosterlitz-Thouless transition: below $T^{\text{BKT}} \sim \rho_\sigma = J_{xx}$ there is a phase of bound vortex-antivortex pairs. 

Thirdly, we could consider alternate gauge field configurations in this two-dimensional system \cite{LKB,HolzmannKrauth}. If the magnetic field is normal to the plane, interspecies currents vanish, while intraspecies currents follow the curl of the gauge field. If the field is uniform, intralayer currents cancel in the bulk but not on the sample boundary. The edge state currents in the two layers are parallel-flowing, giving non-zero density current and zero pseudospin current. Consequently, such edge currents would be observable in the superfluid phase, but not in the Mott phase, unlike the spin-Meissner currents discussed so far. More details on gauge field configurations are offered in the Supplementary Material \cite{supplement}.

Finally, let us note that compared to the energy gap of Eq. (\ref{Eq:Delta_sigma}), the Mott energy scale dominates, $\Delta_\rho \gg \Delta_\sigma$, consistent with our assumptions of strong coupling. There is a distinct regime in which the Mott and phase coherence energy scales are inverted. Previous work on two-leg bosonic ladders has shown that it is possible to achieve the Mott transition at significantly lower energy scales than the phase coherence: $\Delta_\rho \ll \Delta_\sigma$ \cite{OrignacGiamarchi, donohue_giamarchi, Crepin}. This occurs in the regime of weakly coupled chains where $g$ is perturbative compared to all other energy scales. Defining $\theta_{\rho,\sigma} =( \theta_1 \pm \theta_2)/\sqrt{2}$  together with the canonically conjugate $\phi_{\rho,\sigma} = (\phi_1 \pm \phi_2)/\sqrt{2}$, the one-dimensional limit of the system in Eq. (\ref{Eq:H}) reduces to a sum of Luttinger liquid Hamiltonians represented by parameters  
$K_{\rho,\sigma} \sim \sqrt{t/U}(1\pm V_\perp/U)^{-1/2}$ and $u_{\rho,\sigma} = a\sqrt{tU}(1\pm V_\perp/U)^{1/2}$  (plus corresponds to $\rho$) for relatively weak interactions. Additionally, there is a sine-Gordon term of the form $ g \cos(\sqrt{2}\theta_\sigma + a'A_\perp ) \left(1 +  2\cos(\sqrt{8}\phi_\rho )\right)$ \cite{Crepin}. 

Renormalization-group equations show that the $\theta_\sigma$ field becomes gapped first, leading to asymptotically constant correlation functions as in the strongly interacting case. Apart from the difference in parameters, the energy gap below which the correlations have this property is given by Eq. (\ref{Eq:Delta_sigma}) with $1/4K_{\sigma}$ being replaced by $1/2K_{\sigma}$ \cite{OrignacGiamarchi}.
On energy scales below $\Delta_\sigma$, the term in $\cos(\sqrt{8}\phi_\rho)$ remains. The Mott gap takes the form $(g_{\rho}=ga/u_{\rho})$

\begin{equation}
\Delta_\rho \sim \Delta_{\sigma} g_\rho^{ \frac{1}{ 2 - 2 K_\rho } }.
\end{equation}


\noindent Since, as compared to the strongly interacting regime, the two energy scales are inverted, $\Delta_\rho \ll \Delta_\sigma$, observation of the Mott phase along with the Meissner phase requires probing correlators at very low energy scales $\Delta_\rho$. This can be improved by increasing $V_\perp$, which lifts both energy scales $\Delta_\rho$ and $\Delta_\sigma$. This is consistent with our conclusion that $V_\perp$ favors the $ \rho = 1 $ Mott phase, according to the phase diagram of Fig. 1. The introduction of anisotropies drives down the energy scale $\Delta_\sigma$. Such anisotropies can be between hopping terms $t_1\neq t_2$ or intra-species interactions $U_1\neq U_2$. In this sense the isotropic case introduced in Eq. (\ref{Eq:H}) is optimal.

Firstly, the setup presented here has long been possible with Josephson junction arrays \cite{geerligs_et_al,van_der_zant_et_al}. We present such a realization with realistic experimental estimates in the Supplementary Material \cite{supplement}. An early study of the vortex lattice in Josephson-junction arrays, but without considering the Mott transition, has been performed in Ref. \cite{kardar}. In the simplest realization, each species corresponds to a Josephson junction chain. The chains are coupled through a Josephson coupling as well as a visible capacitive interaction and there is a real magnetic field threading the inter-chain plaquettes. The prerequisite of one Cooper pair per rung necessary to access the Mott phase can be achieved through current
technology \cite{KochLeHur}. Another realization of the Hamiltonian of Eq. (\ref{Eq:XXZ}) can be obtained as proposed in Ref. \cite{affleck_garate}, by placing an array of Josephson junctions in the vicinity of a bulk superconductor. The spin degree of freedom then describes total density fluctuations on the superconducting islands. 


Secondly, with cold-atoms a one-dimensional setup is possible \cite{Denschlag}. Recently, staggered artificial gauge fields have been realized \cite{Aidelsburger, Simonet}. Very recently, uniform artificial magnetic fields have been realized \cite{HofstadterCA}: $^{87}\text{Rb}$ atoms have been loaded into tilted optical square lattices; the tilt in one direction suppressed the hopping due to a detuning between neighboring sites. An additional pair of lasers whose detuning was matched to that of the tilt reinstated a complex hopping term, which mimics the Peierls phases acquired by charged particles in a magnetic field. Implementation of a two-leg ladder based on this system requires merely confining the condensate to two columns by use of a parabolic potential. 

In general the on-site interactions dominate $V_\perp \ll U$ \cite{jaksch_bruder_et_al}. Interspecies interaction can be enhanced by the introduction of an additional fermion species that interacts with the bosons $H_f = -\sum_{\langle ij \rangle} t_f f_{i}^\dagger f_j + \text{H.c.}, \;\; H_{\textit{bf}}=V \sum_{\alpha i} n_{\alpha i} n_{f i}.$ Integration of the fermions leaves bosons with repulsive interaction inter-species. This reads $V^2 a/(4\pi|v_F|) \sum_{\alpha\alpha' i} n_{\alpha i} n_{\alpha' i}$ where $\alpha$ denotes species. Longer-range interaction with the fermions induces longer-range interaction between the bosons. Alternatively, the $V_\perp$ is currently realizable with dipolar interactions \cite{lahaye_et_al}. In cold atom experiments, the Mott insulating phase can be probed by measuring local density fluctuations \cite{Esteve} $\langle \rho_i^2 \rangle - \langle \rho_i \rangle^2$. The Meissner phase is characterized by non-vanishing relative density fluctations $\langle \sigma_i^2 \rangle - \langle \sigma_i \rangle^2$. Both can be accessed with \textit{in situ} measurements \cite{in_situ} whereas the total density is locked. Currents can be probed by studying density modulations following anisotropic quenching of the kinetic energy \cite{killi_et_al}. The current response in the spin sector to a magnetic field is given by Eq. (\ref{Eq:jM}), which is the Meissner response. For more details, see the Supplementary Material.

To summarize, it is possible to realize a bosonic insulating phase with a spontaneous and persistent response which directly opposes the magnetic field in a case of a two-component bosonic Hubbard model with total density one.  The associated fluxon quantization in a loop type geometry encodes topological aspects of the spin superfluid. The phase coherence in the Mott insulating regime can be analyzed with current technology in ultracold atoms \cite{Gerbier}. In the strong-field limit where the spin Meissner effect is impossible, we recover the chiral Mott phase with a staggered current pattern found in Ref. \cite{dhar_et_al}. Our analysis could be extended to high-Tc superconductors in the underdoped regime \cite{Bozovic,Fauque,AffleckMarston,Varma,Chak}, and to low-dimensional symmetry protected topological phases \cite{SPTVish, SPTRyu}.

We thank I. Affleck, T. Giamarchi, S. M. Girvin, E. Orignac, A. Paramekanti, Z. Ristivojevic, G. Roux, P. Simon and I. Spielman for fruitful discussions. This work was supported in part by NSF DMR 0803200 and by PALM Labex at Paris-Saclay.

\clearpage
\newpage



This supplement contains details on the mean-field and density-matrix renormalization group phase diagrams; a derivation of the strong-coupling spin-$\frac{1}{2}$ Hamiltonian used in the main text. We also provide a derivation of the renormalization-group equations for the sine-Gordon model and discuss extensions of the model presented in the main text to higher dimensions and a two-dimensional coupled-ladder system. Finally, we discuss possible experimental realizations of the system in Josephson-junction arrays and ultra-cold atoms/molecules.


\section{Mott insulator to superfluid phase transition}

\subsection{Atomic limit}
The atomic limit serves to determine the boundaries of the Mott phase as a function of $\mu$ in the limit of vanishing hopping. For example, we determine here boundaries for the $\rho=1$ Mott lobe. 

Consider the Hamiltonian of Eq. (1) of the main text, suppressing gauge fields. We use the Fock states of an isolated site $|n_1 n_2\rangle$, with $n_{1,2}$ denoting the occupancy of each species.  The atomic limit Hamiltonian for the $\rho=1$ block in the ordered basis $|1,0\rangle,\; |0,1\rangle$ is

\begin{equation}
H(\rho = 1) = \left( \begin{array}{cc} 
-\mu & -g \\
-g & -\mu
\end{array}  \right).
\end{equation}

\noindent The eigenvalues in this block are $-\mu \pm g$. Next, the $\rho=2$ block, in the ordered basis $|2,0\rangle, |1,1\rangle, |0,2\rangle$, is

\begin{equation}
H(\rho = 2) = \left( \begin{array}{ccc} 
U-2\mu & -\sqrt{2}g & 0 \\
-\sqrt{2}g & V_\perp - 2\mu &  -\sqrt{2}g \\
0 & -\sqrt{2} g & U - 2\mu
\end{array}  \right).
\end{equation}

\noindent The eigenvalues in this block are 
\begin{eqnarray}
U - 2\mu, \frac{1}{2} \left(U \pm \sqrt{16g^2 + (U-V_\perp)^2 } + V_\perp-4\mu \right).
\end{eqnarray}

\noindent Varying $\mu$ changes the ground state occupancy. For $U\rightarrow \infty$, as considered in the phase diagram of Figure 1 of the main text, the atomic ground state is the $\rho = 1$ ground state if $V_\perp + g > \mu > -g$; for $\mu < - g$, the ground state is the vacuum; for $\mu > V_\perp + g$, the ground state has $\rho \geq 2$ bosons per site. For $ \mu= - g $ and $ \mu=V_\perp + g $ the atomic ground state is degenerate. Next, we consider the effect of the kinetic term.

\subsection{Phase diagram from mean-field theory}
Consider the Hamiltonian of Eq. (1) in the main text. We obtain the boundary between the Mott phase at density $\rho=1$ and the superfluid phase. The site-factorizable variational wavefunction accounts for fluctuations by $\pm 1$ in the total site density around $\rho=1$: 

\begin{eqnarray}
\label{Eq:VarPsi}
|\psi\rangle = \prod_i &(& n_i |0 0\rangle_i + a_i |1 0\rangle_i + b_i |0 1\rangle_i + \nn \\
 && x_i |2 0\rangle_i + y_i |1 1\rangle_i + z_i |0 2\rangle_i ).
\end{eqnarray}

\noindent $|n_1 n_2\rangle_i$ is the Fock state $i^{\textit{th}}$ site. From here on, we assume that the coefficients are uniform, $a_i=a$ etc., and therefore the normalization condition is

\begin{equation}
|n|^2 + |a|^2  + |b|^2 + |x|^2 + |y|^2 + |z|^2 = 1.
\end{equation}

\noindent These coefficients are determined from the minimization of the variational energy  $\langle \psi | H | \psi \rangle$. The variational energy per site $E_{\textit{var}}$ is composed of

\begin{eqnarray}
E_{\textit{var}} &=& U(|x|^2  + |z|^2)+V_\perp|y|^2 - \mu \langle b_1^\dagger b_1 + b_2^\dagger b_2 \rangle \nonumber \\
&& - g \langle b_1^\dagger b_2 + b_2^\dagger b_1 \rangle - t Z \big( \langle b_1^\dagger \rangle \langle b_1 \rangle + \langle b_2^\dagger \rangle \langle b_2 \rangle \big).
\end{eqnarray}

\noindent The first row and the first term on the second row are obtained from the atomic part of the Hamiltonian. In the kinetic part, $Z=2d$ is the number of near-neighbors on the $d-$dimensional square lattice. The order parameters depend on the coefficients in the variational wavefunction:

\begin{eqnarray}
\langle b_1^\dagger \rangle &=& \sqrt{2}x^* a+y^*b+a^*n, \nonumber \\
\langle b_2^\dagger \rangle &=& y^*a + \sqrt{2}z^*b+ b^*n, \nonumber \\
\langle b_1^\dagger b_2 \rangle &=& a^* b +   \sqrt{2} x^* y +  \sqrt{2} y^* z, \nonumber \\
\langle b_2^\dagger b_1 \rangle &=& b^* a +  \sqrt{2} y^* x +\sqrt{2} z^*y, \nonumber \\
\langle b_1^\dagger b_1 \rangle &=& |a|^2 +  2|x|^2 + |y|^2, \nonumber \\
\langle b_2^\dagger b_2 \rangle &=& |b|^2 +  2|z|^2 + |y|^2.
\end{eqnarray}

\noindent The order parameters determine the two phases of interest, Mott phase at total density $\rho=1$ and superfluid phase, based on the classification in Table \ref{Tab:OP}. In the last row of Table \ref{Tab:OP}, $\langle b_1^\dagger b_1 + b_2^\dagger b_2 \rangle \neq 1$ in the superfluid phase with the exception of a special line that starts at the tip of the Mott lobe. 
In Figure 1 of the main text, only the contour delimiting the region of constant density $\rho=1$ is shown.


\begin{table}
\begin{tabular}{|c|c|c|}
\hline
order parameter & Mott phase $\rho = 1$ & superfluid \\
\hline
$\langle b_{1,2}^\dagger \rangle \neq 0$ & false & true \\
$\langle b_{1}^\dagger b_{2} \rangle \neq 0$     & true  & true \\
$\langle b_{1}^\dagger b_{1} + b_{2}^\dagger b_{2} \rangle = 1$ & true & false \\
\hline 
\end{tabular}
\caption{\label{Tab:OP} Identification of Mott $\rho=1$ and superfluid phases based on values of order parameters.}
\end{table}

In Figure 1 of the main text, we have plotted the mean-field phase diagram of hard core bosons at finite $V_\perp$. The hard core constraint is implemented requiring $|x|=|z|=0$. The boundaries of the Mott lobe $\mu=-g$ and $\mu=V_\perp + g$ are consistent with the calculation in the atomic limit. We remark that Josephson coupling itself is enough for the phase with $\rho=1$ to exist: In the limit $V_\perp \rightarrow 0$, the $ \rho = 1 $ Mott phase is located between $\mu =\pm g$.

\subsection{DMRG phase diagram}
We use the density matrix renormalization group implementation of Ref. \cite{alps} to draw the phase diagram for a one-dimensional system. We consider the Hamiltonian of Eq. (1) of the main text, with gauge fields suppressed. We determine the ground state in a block of fixed particle number (setting $\mu=0$). For $N$ particles on a finite ladder of length $L$ with open boundary conditions, the ``half-filling'' $\rho=1$ block corresponds to $N=L$. 

With respect to the reference ground state energy at $\rho=1$, the particle and hole excitation energies are

\begin{equation}
\label{Eq:MuPM}
\pm\mu^\pm = E( N = L \pm 1 ) - E( N = L ),
\end{equation}

\noindent and the Mott gap is defined as $\Delta_\rho = \mu^+ - \mu^-$. The particle and hole excitation energies determine the boundaries of the Mott lobe. These phase boundaries can be plotted in the $t$ (hopping along the chains) and $\mu$ plane. The Mott phase has a finite gap $\Delta_\rho$. 

\begin{figure}[t!]
\includegraphics[width=0.65\linewidth]{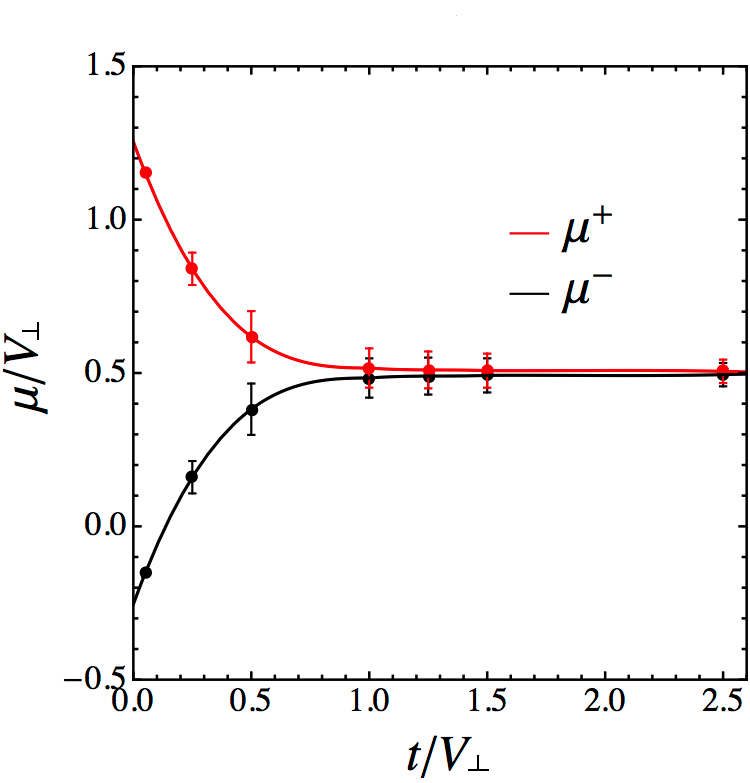}
\caption{\label{Fig:misf-hcb}(Color online) Mott insulator phase for hardcore bosons with $V_\perp = 1.0$ and $g=0.25$. The top and bottom curves represent $\mu^+$ and $\mu^-$, respectively. The error bars are from linear fits of $\mu^\pm (1/L)$. The lines going through the points are guides to the eye. }
\end{figure}

The energy gaps obtained from Eq. (\ref{Eq:MuPM}) are expected to scale with the finite size of the system, $L$ \cite{kuehner_et_al},
\begin{eqnarray}
\mu^{\text{MI}} &\sim& A_0 + A_1 \frac{1}{L} + A_2 \frac{1}{L^2},  \nonumber \\
\mu^{\text{SF}} &\sim& B_0 + B_1 \frac{1}{L}.
\end{eqnarray}
The corrected  phase boundaries $\mu^\pm$ are determined from extrapolations to the thermodynamic limit $1/L \rightarrow \infty$ from a polynomial fit. \noindent In Figure \ref{Fig:misf-hcb} of this supplementary material and in Figure 1 of the main text we show the results for finite-size extrapolation from $L=24,36,48,64,80$. The error bars are obtained from linear fits, and are magnified by a factor of $10^2$. The DMRG routine of Ref. \cite{alps} was run with 2 sweeps, a maximum of 60 states in the truncated space, and 20 warm-up states. 

\section{Derivation of gauged spin $\frac{1}{2}$ Hamiltonian} 
An effective Hamiltonian can be derived in the $\rho=1$ Mott phase to order $\frac{t^2}{U}$ and $\frac{t^2}{V_\perp}$. The derivation is valid in arbitrary spatial dimension. For simplicity, we shall suppress the gauge fields and reintroduce them at the end. The kinetic term for a bond $ij$ is 

\begin{equation}
\label{eq_tij}
T_{ij}  = -t \sum_\alpha ( b_{\alpha,i} ^\dagger b_{\alpha,j}  + h.c. ),
\end{equation} 

\noindent This operator takes half-filled $ \rho = 1 $ states into doubly occupied and empty states. Therefore an effective Hamiltonian can be found at the second order in perturbation theory in $T_{ij}$ (see, for example, \cite{Kuklov}):

\begin{equation}
\label{ha}
(H_\sigma)_{\alpha \beta} = -\sum_{\langle i j \rangle}\sum_\gamma \frac{(T_{ij})_{\alpha \gamma}(T_{ij})_{\gamma \alpha}}{E_\gamma - \frac{1}{2}(E_\alpha + E_\beta)}.
\end{equation}

\noindent Above, $\alpha$ and $\beta$ denote the unperturbed half-filled states,

\begin{eqnarray}
\label{hfbasis}
\alpha:\;  |n_1 n_2 \rangle_i |n_1 n_2 \rangle_j \in &\{& |1 0\rangle_i |1 0\rangle_j, |1 0\rangle_i |0 1\rangle_j,  \nn \\
&&|0 1\rangle_i |1 0\rangle_j, |0 1\rangle_i |0 1\rangle_j \}. \nn
\end{eqnarray}

\noindent The $\gamma$ states are the excited states produced by a single application of the operator $T_{ij}$

\begin{eqnarray}
\gamma:\; |n_1 n_2 \rangle_i |n_1 n_2 \rangle_j \in \{ && |2 0\rangle_i |0 0\rangle_j, |1 1\rangle_i |0 0\rangle_j, |0 2\rangle_i |0 0\rangle_j, \nonumber \\ 
&& |0 0\rangle_i |2 0\rangle_j, |0 0\rangle_i |1 1\rangle_j, |0 0\rangle_i |0 2\rangle_j \}. \nn
\end{eqnarray}

The spin Hamiltonian is now determined by reexpressing the Fock kets $|\uparrow\rangle \equiv|1 0\rangle$ and  $|\downarrow\rangle \equiv|0 1\rangle$. Our original bosons correspond to the Schwinger boson representation of spin: $\sigma_x = b_1^\dagger b_2 + b_2^\dagger b_1$, $\sigma_y = -i b_1^\dagger b_2  +i b_2^\dagger b_1$, and $\sigma_z = b_1^\dagger b_1 - b_2^\dagger b_2$.  In the ordered basis 
$\{|\uparrow\rangle, |\downarrow\rangle \} \otimes \{ |\uparrow\rangle, |\downarrow\rangle \}$, the Hamiltonian reads
\begin{equation}
\label{hxxzpre}
H_\sigma=\left(
\begin{array}{cccc}
  -\frac{4t^2}{U} & 0 & 0 & 0 \\
  0 & -\frac{2t^2}{V_\perp} &  -\frac{2t^2}{V_\perp}  & 0 \\
  0 & -\frac{2t^2}{V_\perp} &  -\frac{2t^2}{V_\perp} & 0 \\
  0 & 0 & 0 & -\frac{4t^2}{U} \\
\end{array}\right),
\end{equation}
or, more compactly in terms of Pauli matrices
\begin{eqnarray}
\label{Eq:XXZ}
&&H_\sigma = -\sum_{\langle i j \rangle} \left( 2 J_{xx} (\sigma^+_i \sigma^-_j + \text{H.c.}) - J_z \sigma^z_i \sigma^z_j \right) -g \sum_i \sigma^x_i, \nn \\
&&J_{xx} = \frac{t^2}{V_\perp},\;\; J_z = t^2\left(\frac{2}{U} - \frac{1}{V_\perp}\right).
\end{eqnarray}
Above, we have used $\sigma^\pm_i = \frac{1}{2} (\sigma^x_i \pm \sigma_i^y)$. The conversion term $g$, or the spin-flip operator, has been added back. It does not create doubly-occupied states out of half-filled states. We have left out the constant term in the Hamiltonian equal to $-t^2\left( 2/U + 1/V_\perp \right)$, which does not affect the dynamics. Consider now adding the gauge fields of Eq. (1) in the main text. Then

\begin{equation}
\label{Eq:HXXZB}
H_\sigma = -\sum_{\langle i j \rangle} \left( 2 J_{xx} (\sigma^+_i \sigma^-_j e^{i\phi_{ij}} + \text{H.c.}) - J_z \sigma^z_i \sigma^z_j \right) -g \sum_i \sigma^x_i.
\end{equation}

\noindent The flux is defined as $\phi_{ij} = a A^1_{ij} - a A^2_{ij} - a' A_{\perp i} + a' A_{\perp j}$. Through a gauge transformation this can be brought to the form of Eq. (5) in the main text.

\section{Renormalization-group equations}

In this section we derive the asymptotic form of the gaps $\Delta_\sigma$ and $\Delta_\rho$ in equations (8) and (10) of the main text.

We consider a generic sine-Gordon Hamiltonian
\begin{equation}
H = \frac{1}{2\pi}\int dx \left( u K (\nabla \theta)^2 + \frac{u}{K}(\nabla \phi)^2  \right) + \frac{g}{a} \int dx \cos\left( \beta \phi \right),
\end{equation}
for which we derive the renormalization group equations for $g$ and $K$ to second order in the perturbation $g$. We will finally use a duality relation to derive the renormalization-group equations for $g\int dx \cos(\beta \theta(x))$. The dimensionless $\beta$ is related to the scaling dimension. In the main text, we have encountered a sine-Gordon term in Eq. (7) which was $\sim \frac{g}{a} \int dx \cos(\theta_\sigma(x))$, with $\beta = 1$; a sine-Gordon term $\cos(\sqrt{8} \phi_\rho(x))$ before Eq. (10), with $\beta = \sqrt{8}$, etc. The introduction of the (artificial) gauge fields does not change the arguments below.

Following Ref. \cite{Giamarchi}, we shall require that the two-point correlation function remains invariant under a change of the low distance cutoff. We expand the interacting theory zero temperature two-point correlation function
\begin{equation}
R(r_1 - r_2) = \langle e^{i \phi(r_1)} e^{- i  \phi(r_2)}\rangle
\end{equation}
to second order in $g$. 

\begin{widetext}

The expansion of the correlation function (equivalently, of the partition function) to second order in the coupling $g$ is
\begin{eqnarray}
\label{Eq:expansion_2nd_order}
R(r_1 - r_2) =  \langle e^{i \phi_1} e^{- i  \phi_2}\rangle_0 
+ \frac{1}{2^3} \left( \frac{g}{u a} \right)^2 \sum_{\epsilon',\epsilon''=\pm 1} \int d^2 r' d^2 r'' \langle e^{i\phi_1} e^{-i \phi_2} e^{i \epsilon' \beta \phi'} e^{-i \epsilon'' \beta \phi''}  \rangle_{0,\text{c}} 
\end{eqnarray}

For brevity, we denote $\phi_1 = \phi(r_1)$ etc. Integrals $\int d^2 r \equiv u \int_0^\infty dx \int_0^\infty d\tau$. The connected correlation function means $\langle e^{i\phi_1} e^{-i \phi_2} e^{i \epsilon' \beta \phi'} e^{-i \epsilon'' \beta \phi''}  \rangle_0 -\langle e^{i\phi_1} e^{-i \phi_2} \rangle_0 \langle e^{i \epsilon' \beta \phi'} e^{-i \epsilon'' \beta \phi''}  \rangle_{0}$.

In the gaussian theory correlation functions of products of exponentials are power-laws:
\begin{equation}
\left\langle \prod_j e^{i A_j \phi_j} \right\rangle_0 = \delta\left(\sum_j A_j\right) e^{-\frac{K}{2} \sum_{i < j} A_i A_j F(r_i - r_j)},
\end{equation}
where $F(r_i - r_j) \equiv \log |r_1-r_2|/a$ and the length $a$ is the small distance cutoff. The simplest of these correlation functions is the two-point correlation function $R_0(r_1-r_2) = \left( a / |r_1 -r_2|\right)^{\frac{K}{2}}.$

The double integral in Eq. (\ref{Eq:expansion_2nd_order}) is dominated by contributions from nearby terms $r' \approx r''$. Expanding in the small parameter $r = r - r'$, we arrive at:

\begin{eqnarray}
R(r_1-r_2) = R_0(r_1 - r_2 )\Big( 1+ \frac{y^2 \beta^2 K^2}{2^5} F(r_1-r_2) \int_{r>a}\frac{dr}{a} \left(\frac{r}{a} \right)^{3-\frac{\beta^2}{2}K} \Big). \nn
\end{eqnarray}
We have introduced the dimensionless coupling constant $y = \frac{g a}{ u }$. Approximating the parenthesis by an exponential function yields
\begin{equation}
R(r_1 - r_2) \approx e^{-\frac{K}{2} F( r_1 - r_2)} e^{\frac{y^2 \beta^2 K^2}{2^5} F(r_1-r_2) \int_{r>a}\frac{dr}{a} \left(\frac{r}{a} \right)^{3-\frac{\beta^2}{2}K}}.
\end{equation} 
\end{widetext}
We express the two-point correlator as 
\begin{equation}
R(r_1 - r_2) = e^{-\frac{K_\textit{eff}}{2} F( r_1 - r_2)}
\end{equation}
with
\begin{equation}
K_{\textit{eff}}(a)  =  K - \frac{\beta^2 y^2 K^2}{2^4}\int_a^\infty \frac{dr}{a}\left( \frac{r}{a} \right)^{3-\frac{\beta^2}{2} K}.
\end{equation}
The renormalization-group equations arise by requiring that $R(r_1-r_2)$, or equivalently $K_\textit{eff}$, be invariant under a change of the low distance cutoff. We may rewrite the equation above as
\begin{eqnarray}
K_{\textit{eff}}(a) = K -  \frac{\beta^2 y^2 K^2}{2^4} \left( \int_a^{a+da} + \int_{a+da}^\infty \right) \frac{dr}{a}\left( \frac{r}{a} \right)^{3-\frac{\beta^2}{2} K}  \nn \\ 
= K - \frac{\beta^2 y^2 K^2}{2^4} \frac{da}{a} -\frac{\beta^2 y^2 K^2 }{ 2^4 } \int_{a+da}^\infty \frac{dr}{a}\left( \frac{r}{a} \right)^{3-\frac{\beta^2}{2} K} + ... \nn
\end{eqnarray}
The ellipse denotes higher order terms in $\frac{da}{a}$. $K_\textit{eff}$ must remain constant with respect to changes in the low energy scale $a \rightarrow  a + da$. The Luttinger parameter $K$ and the coupling $y$ must flow to accomodate these changes:
\begin{equation}
K(a + da) = K(a) - \frac{\beta^2 y^2 K^2}{2^4} \frac{da}{a}.
\end{equation}
The rescaling of the integrand yields the equation for $y$
\begin{equation}
y^2( a + da )  = y^2( a ) \left( \frac{a+da}{a} \right)^{4-\frac{\beta^2}{2}K(a)}.
\end{equation}
Changing variable such that $a(l) = a e^l$ yields the following equations
\begin{eqnarray}
\frac{dK}{dl} = -\frac{\beta^2}{2^4} y^2 K^2, \;\;
\frac{dy}{dl} = \left( 2 -\frac{\beta^2}{4} K \right) y.
\end{eqnarray}
In the weak-coupling limit we approximate $K(l) \approx K(l = 0)$ and the second equation can be integrated to leading order in $y$. To obtain the analogous equations for $\cos (\beta \theta(x))$, one needs to simply map $K \rightarrow K^{-1}$ in all equations.

As the sine-Gordon term flows to strong coupling, the spectrum will acquire a gap $\Delta$, determined as follows. We define the parameter $l^*$ at which $y$ flows to strong coupling:
\begin{equation}
y(l^*) = 1 = \frac{g a}{u} e^{\left(2 - \frac{\beta^2}{4}K \right)l^*}.
\end{equation}  
Then, we use the fact that within our notations the gap is defined as:
\begin{equation}
l^* = \ln \left( \frac{u}{\Delta a}\right)
\end{equation}
The asymptotic form for the gap $\Delta$ then is
\begin{equation}
\Delta \sim \frac{u}{a} y ^ \frac{1}{2 - \frac{\beta^2}{4}K}.
\end{equation}
If the sine-Gordon term was instead $\int dx \cos(\beta \theta)$, then this would be modified by replacing $K \rightarrow K^{-1}$. Note, $\beta=1$ in Eq. (7) of the main text and $\beta=\sqrt{8}$ to find Eq. (10).

\section{Two-dimensional system}
\begin{figure}[t!]
\includegraphics[width=\linewidth]{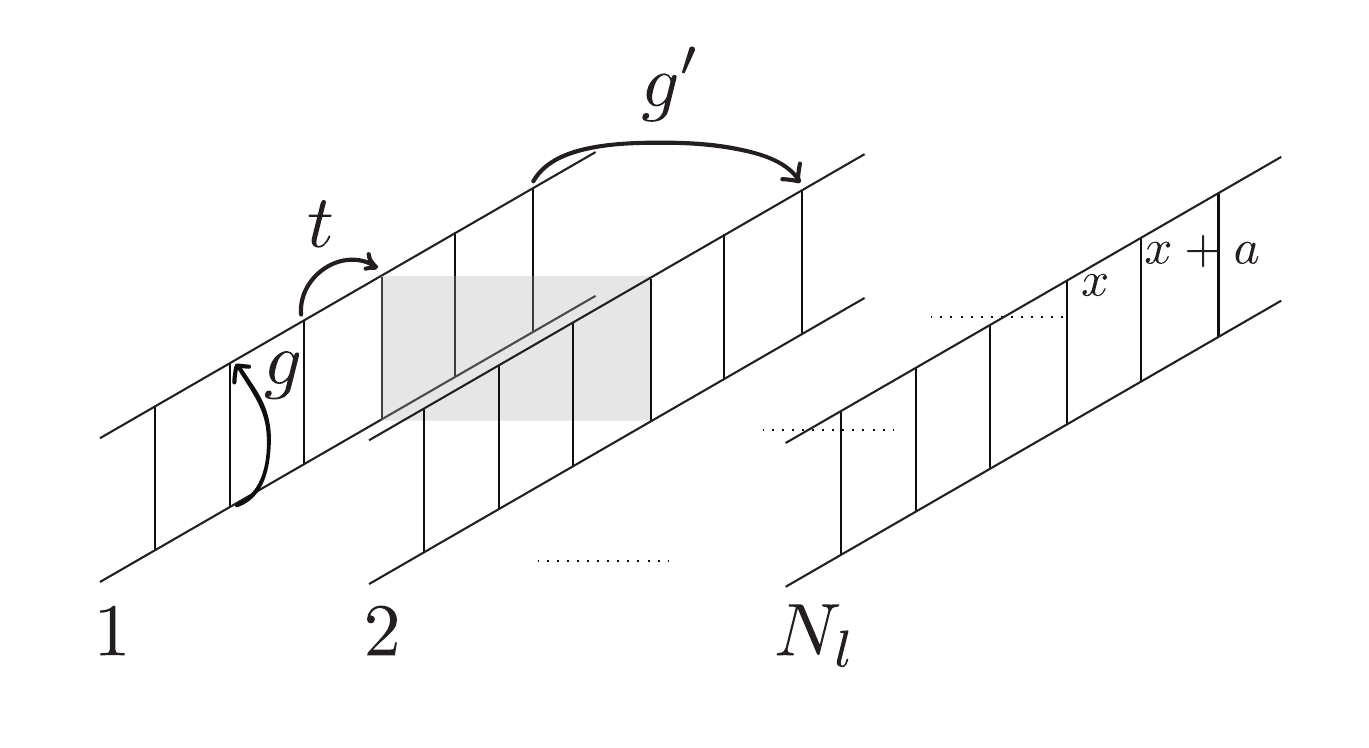}
\caption{\label{Fig:supplement_figure1} Ladder System as a quasi-one-dimensional analogue of the two-flavor two-dimensional system. Ladders $l = 1, ..., N_l$, described by one-dimensional system Hamiltonians $H_l$, are coupled by the nearest-neighbor kinetic term proportional to $g'$. The plaquette around which $\text{phase}_{l,l+1}$ is defined is highlighted in grey. In each ladder, the system is described by a rung Mott insulator with total density 1 and with a spin-Meissner effect between the chains belonging to different planes (flavors). At low-energy the coupling $g'$ will become effective then producing (spin)-coherence in each plane.}
\end{figure}
In this section we construct the Hamiltonian of the two-dimensional system: in the strong-coupling regime, starting from a Gutzwiller Ansatz (Eq. (9) of the main text), and secondly starting from a coupled ladder construction.

\subsection{Strong-coupling regime}

Here, we start from Eq. (5) in the main text using the conventions defined below Eq. (6). Taking $|\psi\rangle = \prod_i (\cos\phi_{\sigma i} |\uparrow \rangle_i  + e^{i\theta_{\sigma i}} \sin \phi_{\sigma i} | \downarrow\rangle_i )$, corresponding to $\langle \sigma_i^+ \rangle = e^{i\theta_\sigma}$, we obtain from the spin Hamiltonian the following variational energy
\begin{eqnarray}
\label{Eq:StrongCoupling2D}
\langle H \rangle = - \sum_{j \in \langle i \rangle } \frac{J_{xx}}{2} \cos( -\theta_{\sigma i} + \theta_{\sigma j} - a ( A_{ij}^1 - A_{ij}^2)) \nn \\ - g\sum_i \cos(\theta_{\sigma i} + a' A_{\perp i}). \nn
\end{eqnarray}
This variational energy corresponds to the saddle point $\phi_\sigma = \frac{\pi}{4}$, corresponding to the $XY$ limit $J_z \rightarrow 0$. In this limit, we expand the $XY$ term $J_{xx}$ in gradients of the field $\theta_\sigma$. On a d-dimensional hypercubic lattice
\begin{eqnarray}
\langle H \rangle &=& \int \frac{d^d x}{a^{d-2}} \frac{J_{xx}}{2} \left( \nabla \theta_\sigma(x) - A^1 (x) + A^2(x) \right)^2 \nn \\
&&-g \int \frac{d^d x}{a^d} \cos(\theta_\sigma(x) - a' A_\perp) \nn \\
&=& \int \frac{d^d x}{a^{d-2}} \frac{J_{xx}}{2} \left( \nabla \theta_\sigma(x) - A^\sigma(x) \right)^2 \nn \\ &&-g \int \frac{d^d x}{a^d} \cos(\theta_\sigma(x) + a' A_\perp). 
\end{eqnarray}

\subsection{Coupled-ladder construction}

In this section we construct a two-dimensional analogue by coupling multiple ladders. The main idea is to visualize each of the two-dimensional plane discussed above as a collection of one-dimensional chains $l=1,....,N_l$ in each plane. The two planes correspond to the two flavors discussed in the main text.  Below, we shall assume that the 
coupling $g'$ between the chains (in each plane) is smaller than than the Josephson coupling $g'$ between the ``layers'' (flavors) such that the effective model at low energy is a collection of coupled ladder systems, where the fixed point of each ladder corresponds to the one-dimensional rung Mott insulator with spin-Meissner currents discussed in the main text.
Below, let $H_l$ be the Hamiltonian corresponding to the $l^{\textit{th}}$ ladder (see Fig. \ref{Fig:supplement_figure1}). Assume that this is the effective Hamiltonian under the energy scales $\Delta_\rho$ and $\Delta_\sigma$, where all ``spin'' fields $\theta_{\sigma,l}$ and all ``charge'' fields $\phi_{\rho,l}$ in each ladder have been gapped.

\begin{figure}[t!]
\includegraphics[width=\linewidth]{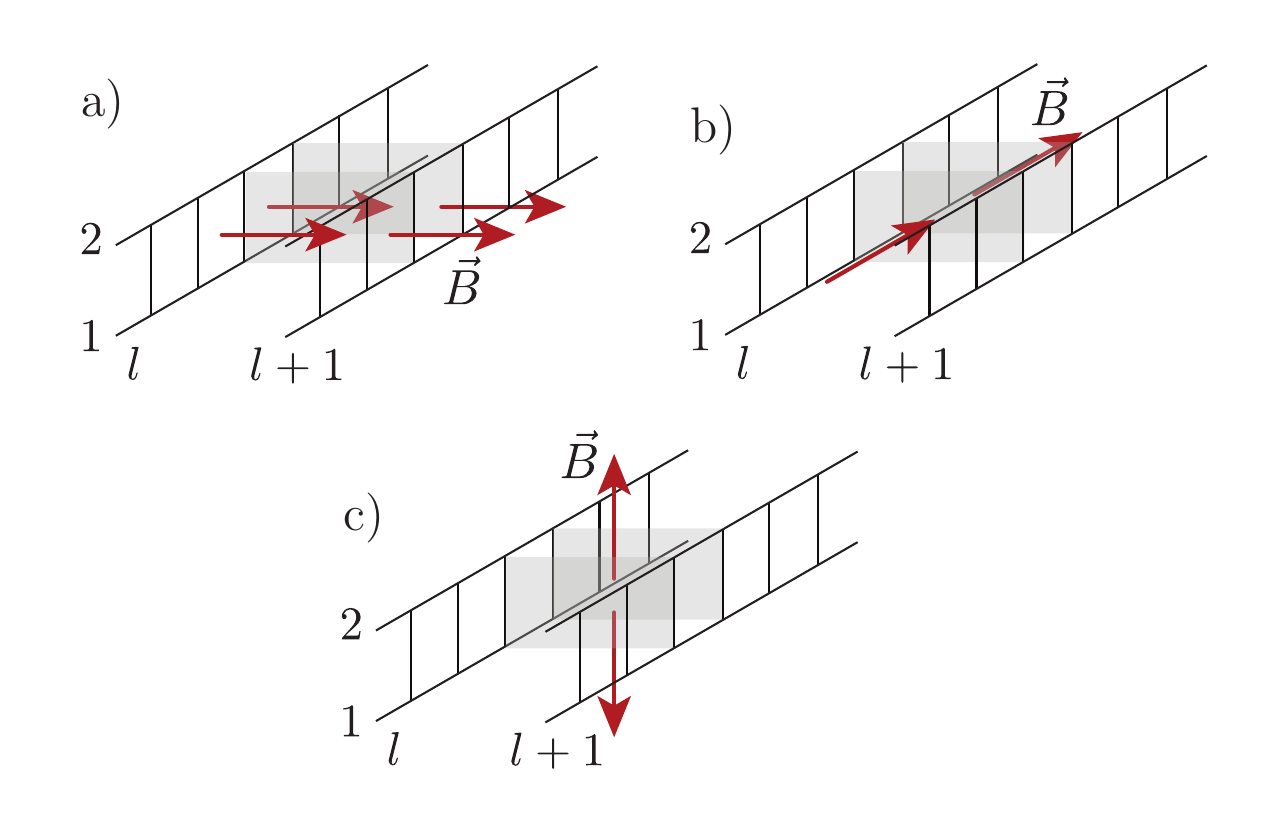}
\caption{\label{Fig:Gauge} Magnetic field configurations in which the gauge field couples only to relative charge $\sigma$. In a) the corresponding gauge field is parallel to the layers and perpendicular to $x$; this is the case studied in the main text with $A_{\parallel\sigma}, A_{\perp} \neq 0$ and $A_{\alpha l}=0$. The ``Meissner'' phase here consists of currents being confined to the layers, with no inter-layer current. In b) the gauge field is parallel to the layers and to the ladders: $A_{\parallel\sigma}=0$, $A_\perp, A_{\alpha l} \neq 0$. One cannot measure a $j_\sigma$ on the ladders, but between ladders. Finally, c) shows the magnetic field emanating from a ``sheet of magnetic monopoles'' situated between the layers: $A_{\parallel\sigma}, A_{\alpha l}\neq 0$, $A_\perp = 0$.}
\end{figure}

Let $g' < \text{min}\left\{ \Delta_{\rho}, \Delta_{\sigma} \right\}$ be the strength of the interladder kinetic terms
\begin{equation}
H_c = -g' n \sum_{l=1}^{N-1} \sum_{\alpha = 1,2} \int dx \cos( \theta_{\alpha,l} - \theta_{\alpha,l+1} + a A_{\alpha l}(x)), 
\end{equation} 
where $n =\frac{1}{2a}$ is the mean linear boson density in chain $l = 1, 2, ... N_{l}$. Hopping out of a chain and into a neighboring chain is only allowed inside of each layer $\alpha = 1,2$. The Peierls phase associated with the kinetic term in layer $\alpha$ is $a A_{\alpha l}(x)$, where $a$ is the spacing between the chains. We recast this in terms of the linear (anti-)symmetric linear combinations, defining $A_{\rho,\sigma l}(x) = \frac{A_{1 l}(x) \pm A_{2 l}(x)}{\sqrt{2}}$ to obtain
\begin{eqnarray}
H_c &=& -2g' n \sum_{l=1}^{N-1} \int dx \cos\frac{1}{\sqrt{2}}( \theta_{\rho,l} - \theta_{\rho,l+1} + aA_{\rho l}) \cdot \nn \\ &&\;\;\;\;\;\;\;\;\;\;\;\;\;\;\;\;\cos \frac{1}{\sqrt{2}}(\theta_{\sigma,l} - \theta_{\sigma,l+1} + a A_{\sigma l})
\end{eqnarray}
Under the energy scale $\Delta_\rho$, $\theta_\rho$ is disordered. Since the term in the previous equation is irrelevant, we expand the Hamiltonian in perturbation theory to second order
\begin{equation}
H^{\textit{eff}}_c \sim - g'_2 n \sum_{l=1}^{N-1} \int dx \cos\sqrt{2}(\theta_{\sigma,l} - \theta_{\sigma,l+1} + a A_{\sigma l}).
\end{equation}
We have let $g'_2=\frac{g'^2}{\text{min}\{\Delta_\rho, \Delta_\sigma \}}$. Under $\Delta_\sigma$, the fields $\theta_{\sigma,l}$ are already pinned to their classical values but up to multiples of $2\pi$. The role of $H_c$ is to impose an additional global phase pinning, on an energy scale of order $(g')^2/\text{min}\{\Delta_\rho, \Delta_\sigma \}$. We note the consistency with the strong-coupling calculation performed in the previous subsection: a gradient expansion here gives a term $\sim -g'_2 (\nabla \theta_\sigma + a A_{\sigma})^2$, that coincides Eq. (\ref{Eq:StrongCoupling2D}) if $g'_2 = \frac{g'^2}{\text{min}\{\Delta_\rho, \Delta_\sigma \}} \sim \frac{t^2}{V_\perp} \sim J_{xx}$.  

The current expectation value between ladder $l$ and $l+1$ can be obtained 
\begin{eqnarray}
\langle j_{\sigma l} (x) \rangle = 2 g_2' n  \big( \sin( a A_{1 l} + \theta_{1l} - \theta_{1,l+1} ) \nn \\ - \sin( a A_{2 l} + \theta_{2l} - \theta_{2,l+1})  \big).
\end{eqnarray}
Recalling now that under $\Delta_\sigma$ the fields $\theta_{\sigma}$ are pinned to their classical values, we have $-\sqrt{2}\theta_{\sigma,l} (x) = a' A_{\perp,l}$, using the definitions of the main text for $A_{\perp,l}$ as the intra-layer gauge field, on chain $l$. Expanding the sines in the limit of weak gauge field, we obtain the following expression for the current,
\begin{eqnarray}
\langle j_{\sigma l} (x) \rangle &\approx& 2 g_2' n  \big(  a A_{1 l} + \theta_{1l} - \theta_{1,l+1} - ( a A_{2 l} + \theta_{2l} - \theta_{2,l+1})  \big) \nn \\
&=& - 2 g_2' n \; \text{phase}_{l,l+1}.
\end{eqnarray}
We have defined the phase 
\begin{equation}
\text{phase}_{l,l+1} = - a A_{1l} + a A_{2 l} + a' A_{\perp l} - a' A_{\perp l+1}
\end{equation}
acquired by the particle around a plaquette between the two layers. We recover the Meissner form of the current for inter-ladder currents in coupled-ladder model presented here.

The extension to finitely many coupled ladders has retained the phase discussed in the main text: the phase is a Mott insulator with total density 1, (charge allowed to fluctuate only along the rung bonds), and a (spin) superfluid in the relative phase $\theta_\sigma$. 

\begin{figure}[t!]
\includegraphics[width=\linewidth]{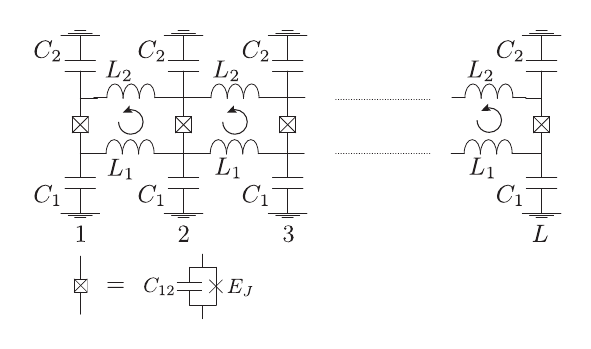}
\caption{\label{Fig:sup_fig_2lljja} Josephson junction based two-leg ladder. Interchain terms are the intra-chain capacitive coupling $C_{12}$ (yielding the $V_{\perp}$ term) and the Josephson term $E_J$ (which gives the $g$ term). Wire junctions can be thought of a superconducting islands. Intra-chain terms are approximated by harmonic terms, hence inductances $L_\alpha$. Capacitive coupling between chains is $C_{12}$. At each site there is a capacitive coupling to ground $C_\alpha$.}
\end{figure}

\subsection{Remarks on gauge field configurations}

In this section we have considered two-dimensional geometries. In the presence of Josephson terms (such as $g \cos(\theta_{\sigma,l}(x) + a' A_{\perp,l}(x))$ in each ladder and $g_2' \cos\sqrt{2}(\theta_{\sigma,l} - \theta_{\sigma,l+1} + a A_{\sigma l})$ between ladders) and in the limit of small field, the (antisymmetric) current $\langle j_{\sigma ij} \rangle = \langle j_{1,ij} - j_{2,ij} \rangle$ was shown to be proportional to minus the phase acquired by the particle around the plaquette defined by the bonds $ij$ in layer 1 and $ij$ in layer 2. A necessary condition for the existence of this (spin) current is that the gauge field couples to the relative charge (and not the total charge: $\langle j_\rho \rangle = 0$ in the Mott phase). The various gauge field configurations that satisfy this condition are enumerated in Figure \ref{Fig:Gauge}. In each case it may be thought of the Meissner currents as circling the plaquettes pierced by the magnetic field vector. 

\section{Experimental Realizations}

\subsection{Josephson-junction two-leg ladder}

The two-leg ladder Hamiltonian presented in the main text can be realized in Josephson junction arrays. Exotic Josephson-junction ladder-type systems have been built experimentally \cite{Bell}. Consider chains $\alpha = 1,2$ consisting of an array of resonators characterized by inductors $L_\alpha$ and capacitors $C_\alpha$. The Hamiltonian for each chain is harmonic \cite{Devoret} 
\begin{eqnarray}
\label{Eq:Halpha}
H_\alpha &=& \sum_{i=1}^L E_{C}^{\alpha} (n_{\alpha i} - n_{\alpha i}^0)^2 \nn \\
         &&+ \sum_{i=1}^{L-1}\frac{E_J^{\alpha}}{2} \left(\theta_{\alpha i} - \theta_{\alpha i+1} + a A_{i,i+1}^\alpha \right)^2 
\end{eqnarray}
$E_C^\alpha$ ($E_{J}^\alpha$) is the charging (Josephson) energy for sites in chain $\alpha$. $L$ denotes the number of  unit cells in each chain (see Fig. 3). 

The offset charge $n_{\alpha i}^0$ is tuned by voltage terms of the form $- V_i^{\alpha} (2e) n_{\alpha i}$ at site $i$.  When the inter-chain capacitance coupling $C_{12}$ (playing the role of the $V_{\perp}$ term in the main text) is sufficiently important, by tuning the offset charges it is possible to attain a state where a single Cooper pair exists in a superposition of states localized on an island on chain 1 and states localized on an island of chain 2, thereby realizing the unit-filling requirement for the Mott phase advertised in the main text. This was shown in the context of a pair of superconducting islands \cite{Bibow,JensKarynJJ}. The second term is the limit of $-\sum_{i=1}^{L-1} E_J^{\alpha} \cos(\theta_{\alpha i} - \theta_{\alpha i+1} + a A_{i,i+1}^\alpha)$ when 
\begin{equation}
\label{Eq:CondAnh}
E_J^\alpha \gg E_C^\alpha,
\end{equation}
such that zero-point fluctuations in the field $\theta$ become negligible and anharmonic terms can be discarded. In Fig. \ref{Fig:sup_fig_2lljja} these terms are represented via inductances $L_\alpha$ and capacitances $C_\alpha$, which are related to the energy scales via
\begin{equation}
E_{C}^\alpha = \frac{(2e)^2}{2 C_{\alpha}},\;\; E_J^\alpha = \frac{1}{L_\alpha} \left( \frac{h}{2e}\right)^2.
\end{equation}
For experimentally accessible values, where $C_\alpha$ is typically in the $\text{pF}$ range and $L_\alpha$ in the $10\; \text{nH}$ range, the ratio $\frac{E_J}{E_C} \sim 10^{4}$ and the value $E_J \sim 10 k_B \text{K}$ satisfy well the conditions of Eq. (\ref{Eq:CondAnh}), and the form of Eq. (\ref{Eq:Halpha}) is valid.

On the other hand, the Hamiltonian that couples the two legs retains the full Josephson term
\begin{eqnarray}
H_c &=& \sum_{i=1}^L \Big( -E^{12}_J \cos( \theta_{1i} - \theta_{2i} + a' A_{\perp,i} ) \nn \\
 && + E^{12}_C (n_{1i} - n_{1i}^0 )( n_{2i} - n_{2i}^0 ) \Big).
\end{eqnarray}
With a Cooper Pair Box \cite{VionEtAl} it is possible to attain $E_{C}^{12} \sim E_{J}^{12} \sim 1 \; k_B \text{K}$. More specifically, $E_C^{12} \sim (2e)^2 / C_{12}$, and $E_J^{12} \sim \frac{h}{(2e)^2} G_N \Delta$ \cite{Devoret}, where $G_N$ is the normal state conductance, and $\Delta$ is the superconducting gap, which is typically on the order of a few Kelvins.  For ratios $E_J / E_C \sim 1$ one must retain the full anharmonic term. 

Returning now to the notation of the original Hamiltonian (Eq. (1) of the main text),
\begin{equation}
t \sim E_{J}^{\alpha},\;U = E_C^\alpha, \; g = E_J^{12},\; V_\perp = E_{C}^{12}.
\end{equation}

The strong-coupling regime discussed in the text $t \ll U, V_\perp$ is equivalent to $E_{J}^\alpha \ll E_{C}^{\alpha}, E_{C}^{12}$ and can be attained by suitably increasing the inductances $L_\alpha$.  

The weakly-coupled ladder regime in which $g$ is perturbative requires $E_{J}^{12} \sim E_{C}^{12}$ (for the cosine potential) and $E_{J}^{12} \sim E_{C}^{12} \ll E_{J}^\alpha$ (weak coupling). Note that the charging energy $E_C^{\alpha}=E_C$ just affects the Luttinger exponent of the theory. The Luttinger exponent in each chain (layer) is very large if $E_C^{\alpha}$ is negligible and the Tonks limit would be achieved in a limit where the intra-chain charging energy would formally become infinite. 

\subsection{Ultracold Atoms and Molecules}
In this subsection we discuss connections to recent experimental realizations of synthetic gauge fields in ultracold atoms, as well as means to realize interactions. 

Experimental proposals involving driven or shaken optical lattices \cite{Simonet, Aidelsburger} have achieved staggered gauge field configurations. In particular, Ref. \cite{Aidelsburger} involves the creation of a two-dimensional optical lattice in which along the $x$ direction a superlattice is added which achieves the detuning of every other site in the $x$ direction by an energy offset $\delta$. For $\delta$ much larger than the $x$ kinetic energy scale, the kinetic term is suppressed; kinetic terms with complex hoppings are restored by a pair of appropriately matched lasers (photon-assisted tunneling). A staggered flux configuration along the $x$ direction (and uniform along $y$) is obtained. In particular, a flux of $\pi$ can be achieved, which would allow for the demonstration of the $\chi = \pi$ vortex lattice discussed in the main text (see Figure 1 of the main text). Moreover, uniform field configurations have been realized very recently \cite{HofstadterCA} by replacing the superlattice potential with a tilt, which would allow probing the $\chi < \pi$ region of our phase diagram, and in particular the low-field Meissner phase. The schemes enumerated for synthetic gauge fields rely on temporal modulation of the lattice and not on internal degrees of freedom.

An enhanced interaction $V_\perp$ can be realized with optical lattices of cold dipolar molecules \cite{lahaye_et_al}. The dipolar interaction takes the form
\begin{equation}
U(r) = C \frac{1-3 \cos^2 \theta}{r^3}.
\end{equation}
The distance between the molecules is $r$ and $\theta$ is the angle between their orientations. Returning to our two-leg ladder implementation, the orientation of the molecules controls the relative strength of intra-leg and inter-leg interactions.  As argued in the main text, a large $V_\perp$ is essential for stabilizing the Mott phase with $\rho = 1$.

\end{document}